\documentclass[prb,twocolumn,amsmath,amssymb,floatfix]{revtex4}
\usepackage{graphicx}

\newcommand{\be}{\begin{eqnarray}}
\newcommand{\ee}{\end{eqnarray}}
\newcommand{\bfk}{{\bf k}}
\newcommand{\bfp}{{\bf p}}

\newcommand{\bfQ}{{\bf Q}}

\newcommand{\hf}{\hat{f}}
\newcommand{\hb}{\hat{b}}
\newcommand{\hrho}{\hat{\rho}}

\begin{document}
\draft

\title{Engineering Superfluidity in Bose-Fermi Mixtures of Ultracold Atoms }

\author{D.-W. Wang$^{1,2}$, M.D. Lukin$^1$, and E. Demler$^1$}

\address{$^1$Physics Department, Harvard University, Cambridge, MA 02138
\\
$^2$Department of Physics, National Tsing-Hua University, 
Hsinchu, Taiwan 300, ROC
}
\date{\today}

\begin{abstract}
We investigate  many-body phase diagrams of 
atomic  boson-fermion mixtures loaded in the two-dimensional optical lattice.  Bosons mediate an attractive, finite-range interaction between fermions, leading to fermion pairing phases 
of different orbital symmetries. Specifically,  we show that by properly tuning atomic and lattice parameters 
it is possible to create superfluids with $s$-, $p$-, and $d$-wave pairing symmetry as well as   spin and charge density wave phases. These 
phases and their stability  are analysed within the mean-field approximation for systems of
$^{40}$K-$^{87}$Rb and $^{40}$K-$^{23}$Na mixtures.
For the experimentally accessible regime of parameters, superfluids with unconventional fermion pairing have transition temperature around  a percent of the Fermi energy.
\end{abstract}


\maketitle
Mixtures of quantum degenerate  atoms recently  became a subject of 
intense studies. Examples include recent experimental observations
of instabilities in boson-fermion mixtures \cite{collapse}, 
$s$-wave pairing Superfluidity \cite{chin} and 
condensation of molecules in fermionic mixtures \cite{molecule}. 
Many other intriguing many-body effects have been proposed theoretically. 
They include formation of composite particles \cite{composite}, appearance 
of charge density wave order \cite{burnett}, phonon-induced fermion 
pairing \cite{phonon_interaction,pairing_lattice}, 
and polaronic effects \cite{ludwig}.
In this Letter, we study quantum phases of boson-fermion mixtures 
(BFM) in two-dimensional (2D) optical lattices. We show 
that a number of very interesting many-body phases can be 
observed in such systems by appropriately choosing atomic 
and lattice parameters. These  include charge and spin density 
wave phases (CDW/SDW) as well as superfluid  states with 
unconventional pairing of fermions.
Experimental realization of such systems should provide 
critical insights  into understanding several important 
strongly correlated electron systems, including  
quasi-2D unconventional superconductors, such as 
high $T_c$ cuperates \cite{cuperate} and organic conductors 
\cite{Bechgaard} displaying 
$d$-wave superconductivity, as well as ruthenates \cite{rutheuates} 
and Bechgaard salts \cite{Bechgaard} displaying $p-$wave superconductivity.
\begin{figure}
\includegraphics[width=7cm]{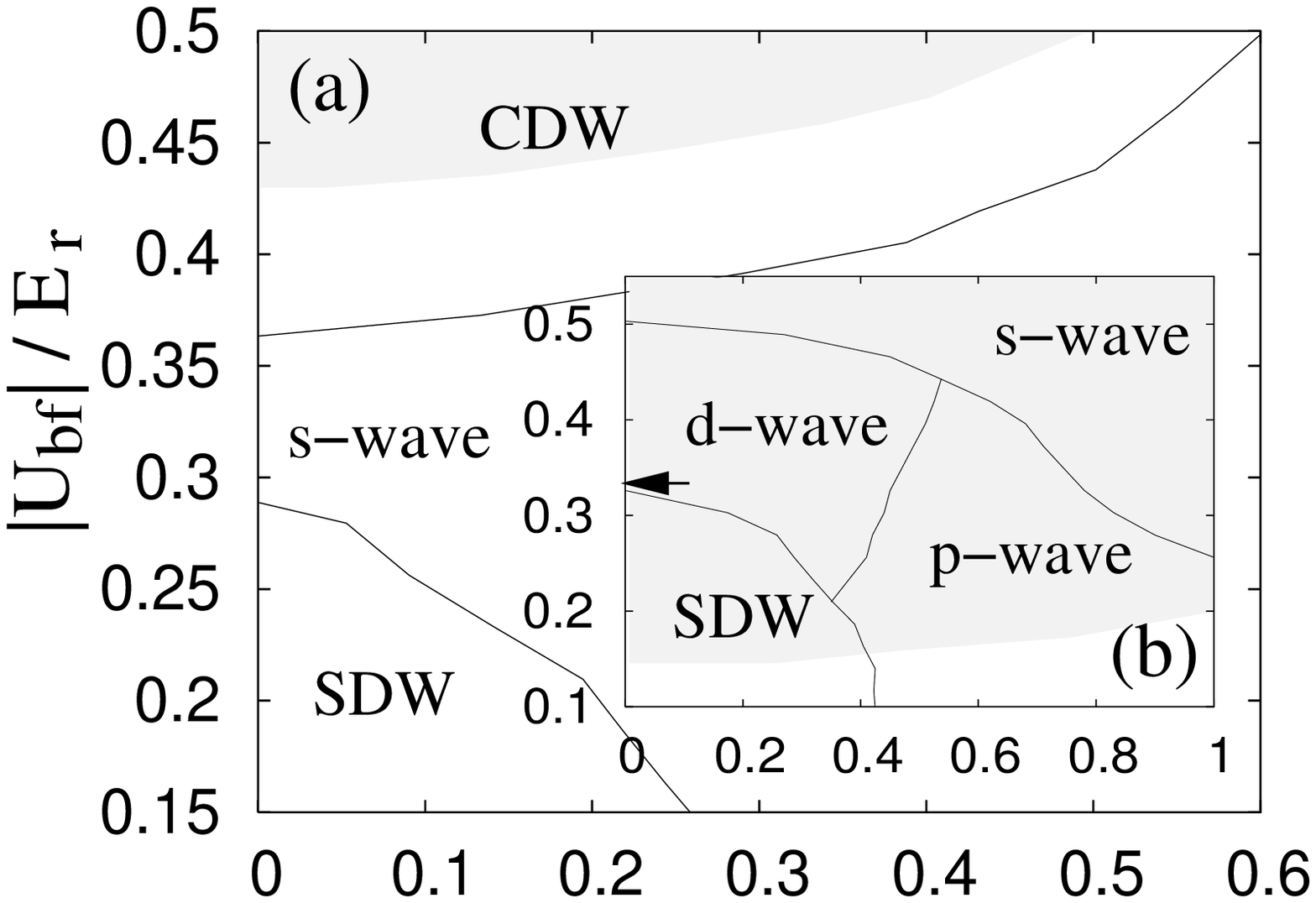}
\includegraphics[width=7cm]{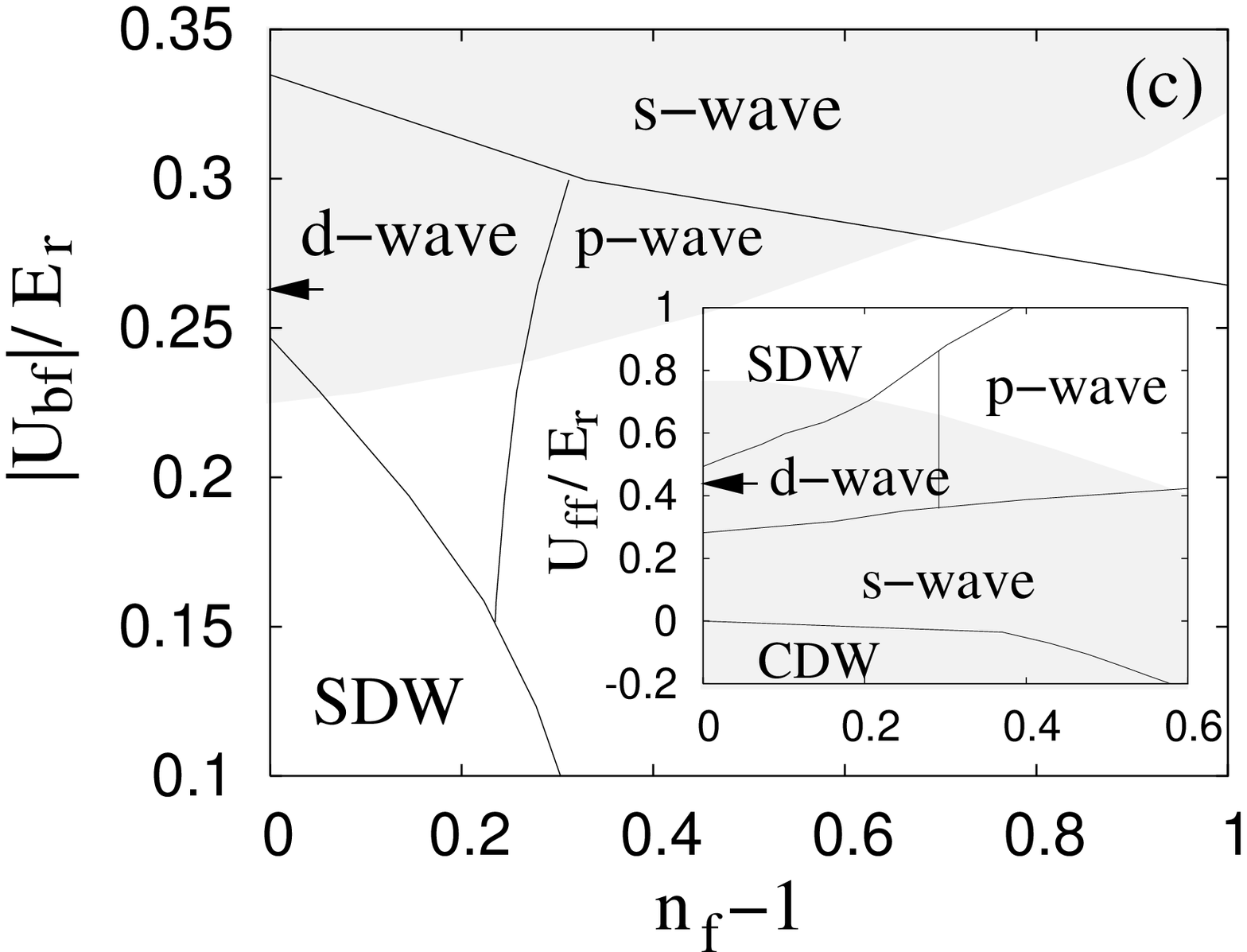}
\caption{
Phase diagrams of (a)-(b) $^{40}$K-$^{87}$Rb mixtures loaded in 2D optical 
lattice with wavelength $\lambda=1060$ and 765.5 nm (blue detuning
from the $D_2$ line of potassium by $\Delta\nu=0.48$ THz) respectively, and
of (c) $^{40}$K-$^{23}$Na mixtures with wavelength $\lambda=1060$ nm. 
We assume that $^{40}$K are prepared in the hyperfine 
states $|F,m_F\rangle=|9/2,-9/2\rangle$ 
and $|9/2,-7/2\rangle$ corresponding to  the pseudo-spin up and down fermions. Bosonic atoms 
$^{87}$Rb and $^{23}$Na are assumed to be trapped in state $|1,1\rangle$. 
$n_f$ is fermion filling fraction and $U_{bf}/U_{ff}$ is
boson-fermion/fermion-fermion onsite interaction.
The in-plane laser intensity is chosen to be $V_0=5 E_r$ for $^{40}$K
and it is $V_{0,z}=30 E_r$ in the $z$ direction
to form a (quasi-)2D systems.
$E_r$ is the recoil energy of $^{40}$K and its value is 
different in the different systems. We choose boson filling 
fraction $n_b=1$ for 
systems A and B but $n_b=9$ for system C respectively. 
The shaded area are the regimes of collapse calculated within the mean 
field approximation. The phase diagrams for hole doping is the
same except for the minor difference in the regimes of collapse.
Arrows indicate the interaction strength away from Feshbach resonance 
\cite{scattering_constant} (it is at $0.84$ $E_r$ and not shown in (a)).
The different lattice strengths for sodium and rubidium 
due to different  optical transition frequencies are included
in the calculation.}
\label{Uff_n_V5_2}
\end{figure}

In the BFM that we consider here, the fermionic atoms are prepared 
as a mixture of  two hyperfine spin states. These two kinds of 
fermionic atoms interact via  short-range {\it repulsive} interaction.
Density fluctuations in a condensate of bosonic atoms induce an attractive 
interaction between fermions,  which is of {\it finite-range}.
Competition between these two types of interactions results in
several many-body phases of fermions in different parameter
regimes.  We show that these phases can be realized for realistic atomic 
systems by tuning the densities of  bosons and fermions,
lattice parameters, or interaction strength via Feshbach resonance. 
We will discuss how systems can be prepared and how 
these phases can be observed in experiments \cite{note}.

Figure \ref{Uff_n_V5_2}  presents  examples 
of mean-field analysis for 
three realistic systems that we considered (these are labelled by systems  A, B, and C respectively in this Letter).
Systems A and B are chosen to correspond to  $^{40}$K-$^{87}$Rb mixtures trapped in optical
lattices of two different laser wavelengths. System A corresponds to 
atoms confined by far-off resonant Nd:YAG laser.  System B   corresponds to  blue-detuned optical lattice tuned closer to the optical resonance of K than that of Rb. Finally system C is $^{40}$K-$^{23}$Na mixture loaded in the  Nd:YAG lattice. In order to show the main features of different competing phases
we use fermion filling fraction and boson-fermion onsite interaction ($U_{bf}$) as a tunable
parameter  (fermion-fermion onsite interaction ($U_{ff}$) is used for the inset of Fig. \ref{Uff_n_V5_2}(c)). 
Interesting quantum phases are
typically obtained by reducing the $U_{bf}$ from its background value.
This can be achieved by either relative shift of bosonic and fermionic lattice \cite{Mandel03}  or   
by tuning the system in the vicinity of Feshbach resonance 
\cite{feshbach}. Other interaction strengths are based on the background 
$s$-wave scattering lengths \cite{scattering_constant}.

Several important results presented in Fig. \ref{Uff_n_V5_2} should be noted. First  of all,  SDW and CDW are dominant near half-filling region due
to the nesting effects. Strong $U_{bf}$ favors CDW (or 
suppresses SDW) phase as shown in (a)-(c), while strong
$U_{ff}$ favors SDW (or suppresses CDW) phase as shown
in the inset of (c). 
This is due to the induced attractive interaction
(proportional to $|U_{bf}|^2$) competing with the onsite repulsion 
between fermions. Second,  the two density wave phases are separated by the superfluid phases. Whereas it corresponds  mostly to the $s$-wave pairing for system A , $d$- and $p$-wave
pairing phases dominate for the  systems B and C. 
Physically, the 
unusual pairing mechanisms arise  when the boson correlation (healing)  length is comparable to lattice constant  resulting in  strong inter-site correlations.
In a typical $^{40}$K-$^{87}$Rb system shown in (a),
such requirement cannot be satisfied because the $^{87}$Rb 
is relatively heavy. However, the ratio of the 
effective mass of bosons to fermions in optical lattice can be reduced 
by tuning the laser frequency closer to the resonance energy of $^{40}$K 
atom  and simultaneously reducing the laser intensity (system B) 
or by simply using a lighter bosonic atoms (system C).
Fig.  \ref{Uff_n_V5_2}(b)-(c) show that both of these two systems 
have wide regime for unconventional ($p$- and $d$-wave) fermion pairing phases.
Finally, within our meanfield analysis, 
a part of these unconventional pairing phases falls in the 
regime of collapse. However, it should be noted that it has been 
shown \cite{lda} that the meanfield approximation strongly overestimates 
the collapse region, while correlation effects beyond the meanfield 
approximation often stabilize  the uniform phases. Thus
Fig. \ref{Uff_n_V5_2}(b)-(c) demonstrate that there 
is a wide range of parameters for which one can observe 
unconventional fermion pairing in BFM systems in 2D optical lattice.

The phase boundaries are
determined by comparing the mean field $T_c$ of these many-body phases.
From our numerical calculation,
$T_c$ for the $p$-wave pairing phase of system B 
is about 0.2-2\% of the Fermi energy (outside
the collapse regime), and it is about 1-3\% for the $d$-wave
pairing phase of system C. We note that our calculations of transition
temperatures rely on similar approximations as employed in analysing
electron pairing in solid state systems with electron-phonon
interactions\cite{Schrieffer}. 

We next detail the microscopic theory resulting in the 
above phase diagrams. When the optical lattice potential is strong enough, 
the BFM system can be described by
the Hubbard type Hamiltonian 
\cite{ludwig,jaksch,walter}:
\begin{eqnarray}
H &=&\sum_{\bfk} \left[
\bar{\epsilon}^b_\bfk \hb_\bfk^{\dagger}\hb_\bfk
+ \bar{\epsilon}^f_{\bfk,\uparrow} 
\hf_{\bfk,\uparrow}^{\dagger}\hf_{\bfk,\uparrow}
+ \bar{\epsilon}^f_{\bfk,\downarrow} \hf_{\bfk,\downarrow}^{\dagger}
\hf_{\bfk,\downarrow}
\right]
\nonumber\\
&&+\frac{1}{\Omega}\sum_\bfk\left[
\frac{U_{bb}}{2}\hrho^b_\bfk \hrho^b_{-\bfk}
+U_{bf}\hrho^b_{\bfk} \hrho^f_{-\bfk}
+U_{ff}\hrho^f_{\bfk,\uparrow} \hrho^f_{-\bfk,\downarrow}\right]
\nonumber\\
\label{H_tot_spinful}
\end{eqnarray}
where $\hb_\bfk$ and $\hf_{\bfk,s}$ are the annihilation operators for bosonic and fermionic atoms with momentum $\bfk$.
$\hrho^b_\bfk=\sum_\bfp \hb^\dagger_{\bfp+\bfk}\hb^{}_\bfp$
is the boson density operators, and 
$\bar{\epsilon}^{b}_\bfk\equiv\epsilon^{b}_\bfk
-\mu_{b}$, where $\epsilon^{b}_\bfk=-t_{b}\gamma_\bfk$ 
is the single particle energy with $t_{b}$ being the tunnelling 
amplitude of bosons between neighboring sites and
$\gamma_\bfk\equiv2\left(\cos k_x+\cos k_y\right)$ (lattice
constant is set to be unit). $\mu_{b}$ is the boson 
chemical potential. Similar notations also apply to fermions with superscript
$f$ and $\hrho^f_\bfk\equiv\hrho^f_{\bfk,\uparrow}
+\hrho^f_{\bfk,\downarrow}$ in Eq. (\ref{H_tot_spinful}).
$U_{bb}$, $U_{bf}$, and $U_{ff}$ are respectively boson-boson,
boson-fermion and fermion-fermion onsite interaction energy, which can be
calculated easily from the $s$-wave scattering length and
the lattice potential\cite{scattering_constant,jaksch,walter}. 
$\Omega$ is the system volume.
For simplicity we do not include the global trapping potential and 
consider systems with uniform densities.

We are interested in the low temperature regime where 
the bosonic atoms are in condensed state ($\mu_b=\epsilon^{b}_0$). Using Bogoliubov approximation, 
one can obtain an effective fermion-phonon coupling Hamiltonian 
\cite{phonon_interaction,bec_book}.
It is well-known \cite{phonon_interaction} that 
the phonon field can be integrated out to provide an effective 
attractive interaction between fermion atoms and hence cause
the fermion pairing. If the phonon velocity $c$ 
is larger than the Fermi velocity $v_f$
(i.e. in the fast phonon limit), the resulting interaction between fermions is 
instantaneous and given by $V_{\rm ind}(\bfk)=-\tilde{V}(1+\xi^2(4-\gamma_\bfk))$,
where $\tilde{V}\equiv U_{bf}^2/U_b$ is the strength of the phonon-induced
attractive interaction and 
$\xi=\sqrt{t_b/2n_bU_{bb}}$ is the correlation(healing) length 
of the bosonic field. 
Such anti-adiabatic limit definitely applies for the system C ($c/v_f
\sim 5$) but should be carefully examined for systems A and B ($c/v_f \sim 1$).
A common approach to including retardation effects is to introduce the
energy cut-off \cite{Schrieffer} in the self-consistent gap equation
(see Eq. (\ref{Delta_k}) below).  Such cut-off may change a prefactor
in the BCS expression for $T_c$ from the Fermi energy, $E_f$, to some
characteristic bosonic frequency. For systems A and B, bosonic
frequencies are not very different than $E_f$. Hence,
retardation effects are small and may be safely neglected. Thus in all cases,
the effective interaction between fermions may be taken as
\begin{eqnarray}
H_{\rm eff} & = &\sum_{\bfk,s}\bar{\epsilon}_\bfk^f
\hf^\dagger_{\bfk,s}\hf^{}_{\bfk,s}+\frac{1}{2\Omega}\sum_{\bfk,s,s'}
V_{\rm eff}^{s,s'}\hrho_{\bfk,s}\hrho_{-\bfk,s'},
\label{H_tot_eff}
\end{eqnarray}
where $V_{\rm eff}^{s,s'}(\bfk)\equiv U_{ff}\delta_{s,-s'}+V_{\rm ind}(\bfk)$.

Following the early work of Micnas {\it et. al.} 
\cite{meanfield_d}, we apply the meanfield approximation
to calculate the $T_c$ of fermion pairing phases 
and that of the competing SDW/CDW phases.
For the superfluid  states, the single particle excitation energy, $E_\bfk$,
has a gap at Fermi surface:
$E_\bfk = \sqrt{(\bar{\epsilon}^f_\bfk+\Sigma_\bfk)^2
+|\Delta_\bfk^{s,s'}|^2}$,
where the gap function, $\Delta_\bfk^{s,s'}$,
is determined by
\be
\Delta_\bfk^{s,s'} &=& \frac{-1}{2\Omega}\sum_\bfp
V_{\rm eff}^{s,s'} (\bfk-\bfp)
\frac{\Delta_\bfp^{s,s'}}{E_\bfp}
\tanh\left(\frac{E_\bfp}{2T}\right).
\label{Delta_k}
\ee
Here $\Sigma_\bfk$
is fermion exchange self-energy within Hartree-Fock (HF) approximation 
and we dropped the spin index due to the spin symmetry.
Finally the fermion chemical potential is determined by 
the known total density of fermions:
\be
n_f&=&\frac{1}{\Omega}\sum_{\bfp,s}\langle \hf^\dagger_{\bfp,s} 
\hf^{}_{\bfp,s}\rangle=
1-\frac{1}{\Omega}\sum_\bfp \frac{\bar{\epsilon}_\bfp^f}{E_\bfp}
\tanh\left(\frac{E_\bfp}{2T}\right).
\label{n_pairing}
\ee
To analyze the results of different gap symmetries, we consider the
following ansatz for the gap function \cite{meanfield_d}:
$\Delta_\bfk^{s,s'}=\delta_{s,-s'}(\Delta_{s0}+\Delta_{s1}\gamma_\bfk
+\Delta_d\eta_\bfk)+\delta_{s,s'}\Delta_p\sin k_x$, where
$\eta_\bfk\equiv 2(\cos k_x-\cos k_y)$. Here $\Delta_{s0}$ and
$\Delta_{s1}$ are for the onsite and extended $s$-wave pairing phase,
while $\Delta_{p/d}$ is for the $p$-/$d$-wave pairing phase.  The
transition temperature $T_c$ is then numerically solved by setting
$\Delta_{s0,s1,d,p}^{s,s'}\to 0$ in
Eqs. (\ref{Delta_k})-(\ref{n_pairing}).  Similar approach is used to
analyze the SDW/CDW phases \cite{unpublished}.  Away from
half-filling, however, the CDW/SDW phases for the commensurate wavevector,
$\bfQ=(\pi,\pi)$, may be less favorable than density wave phases at
incommensurate wavevectors \cite{incomm_SDW}, which are much 
more difficult to analyze than their
commensurate analogues. In this Letter we only discuss SDW/CDW
phases at $(\pi,\pi)$ for simplicity, 
but we note that they provide an accurate
estimate for the regime of general SDW/CDW phases.

\begin{figure}
\includegraphics[width=7cm]{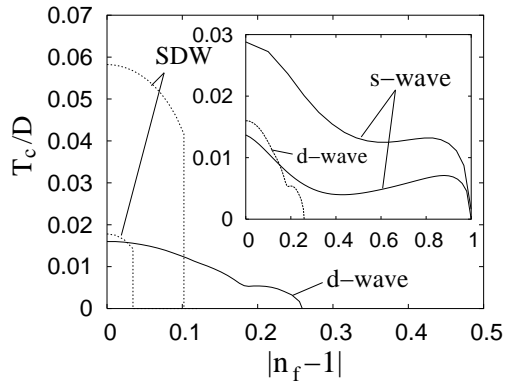}
\caption{Typical $T_c$ of SDW phase and $d$-wave
pairing phases as a function of filling fraction for system C. 
$U_{ff}/E_r=0.52$ and 0.6 for the lower and upper curves of SDW phase 
respectively. $D=4t_f$ is half of the band width.
Inset: same but with $U_{ff}/E_r=0.32$ and 0.28
for the lower and upper curves of the $s$-wave pairing phase 
respectively. $T_c$ of $p$-wave pairing phase is too small and not shown for 
simplicity.
}
\label{Tc_V5_Uff}
\end{figure}

Finally we study the stability of the resulting phases  
\cite{lda,bec_book} by first calculating the 
ground state energy in the mean 
field approximation:
\be
{E}_{MF} &=&\sum_{\bfk}
\left(\bar{\epsilon}_\bfk^f+\Sigma_\bfk-E_\bfk\right)+\mu_f N_{f}
\nonumber\\
&&+\frac{N_f^2}{4\Omega}\left(U_{ff}+2V_{\rm ph}(0)\right)
+\sum_{\bfk,s,s'}\frac{|\Delta_\bfk^{s,s'}|^2}{4E_\bfk}.
\nonumber
\label{E_MF}
\ee
The stability condition then results from requiring
that the compressibility (or bulk modulus, $B$) to be positive. 
At zero temperature it is defined by
$B\equiv -\Omega\left(\partial P/\partial \Omega\right)_{N_f}=
n_f^2\partial^2{\cal E}_{MF}/\partial n_f^2$,
where $P\equiv-\left(\partial E_{MF}/\partial\Omega\right)_{N_f}$ 
is the pressure of the atomic gas and ${\cal E}_{MF}=E_{MF}/\Omega$ 
is the energy density.
We note that the above definition of stability (i.e. $B>0$) is 
different from the previous mean field theories \cite{bec_book}
due to the additional gap energy of the pairing ground state.
This condition is evaluated numerically in our calculation
and the results have been shown
in Fig. \ref{Uff_n_V5_2}. Similar analysis for CDW and SDW phases are 
also presented. It is important to emphasize that the 
mean field results for the 
onset of collapse regime is overestimated due to the lack of correlation 
energy \cite{lda}. More sophisticated self-consistent calculation 
are generally needed for a better quantitative estimate. 

In Fig. \ref{Tc_V5_Uff}, we show the typical $T_c$ of the 
SDW and $d$-wave pairing phases for system C with two different
values of $U_{ff}$. For
strong onsite repulsive interaction SDW phase is favored near the 
half-filling ($n_f=1$) regime,
while $d$-wave pairing become dominant only when $n_f$ is further
away from half-filling. 
In the inset we show the same calculation but with smaller
$U_{ff}$, where the $s$-wave pairing phase become dominant (see
Fig. \ref{Uff_n_V5_2}(c)).
The $T_c$ of $s$-wave pairing phase has a dip at $|n_f-1|\sim 0.5$
because
the onsite and the extended(inter-site) components of $s$-wave pairing
are dominant near half-filling and away from half-filling respectively.
For a typical value of $T_c \sim 0.01-0.03 D$, the
effective BCS coupling strength $\lambda_{BCS}$ is estimated 
to be $0.3-0.45$. 
Therefore most of our results  still fits reasonably 
well into a weak coupling regimes where the meanfield 
approximation is meaningful.

\begin{figure}
\includegraphics[width=7cm]{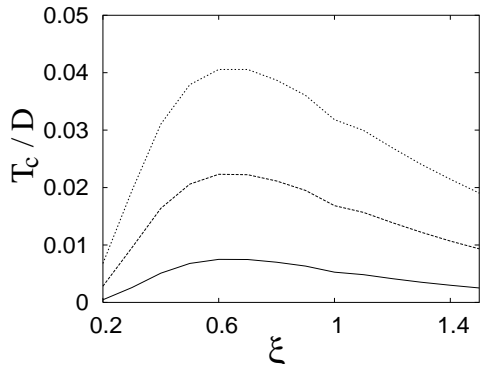}
\caption{$T_c$ of $d$-wave pairing phase at half-filling ($n_f=1$)
as a function of the healing length, $\xi$, which can be easily changed by 
tuning the density of bosonic atoms.
Curves from bottom to up are for $\tilde{V}/E_r=0.53$, 0.88 and 1.23 
respectively, and all the other parameters are the same as used for system C.
}
\label{Tc_xi}
\end{figure}
In Fig. \ref{Tc_xi} we show the $T_c$ of $d$-wave 
pairing phase for system C at half-filling case as a function of $\xi$,
which can be easily tuned by changing the density of bosonic atoms. We find that for different values of $\tilde{V}$, 
the maximum $T_c$ correspond to
$\xi_c\sim 0.6$ (in the units of lattice constant).
For system C, this corresponds to $n_b\sim 9$. 
This observation explains why unconventional pairing phases 
are not present in system A,
which has a very short healing length ($\xi\sim 0.12$) and, 
hence, no inter-site
coupling.

Low-temperature states of a BFM in the 2D optical lattice can be prepared
by a process in which the sympathetic cooling of the 
atomic mixture is followed
by an adiabatic cooling via increasing the lattice potential
\cite{walter} until the phase transition is reached. Here we describe
an altentive method to reach the low temperature unconventional
pairing phases. Consider
the situation in which a condensate mixture of bosonic atoms and
molecules is created. (As in recent experiments \cite{molecule}
molecular BEC is
composed from the two-species fermions in the repulsive interaction side
of Feshbach resonance.) The lattice potential is then turned on and
adiabatically increased such that the molecules are driven into
a strongly localized state (this may not be a Mott insulator state since
the  filling fraction can be less than unity for molecules).  A two-photon
Raman pulse \cite{stirap,melting} is then  applied to dissociate the bound molecule
states into a two-atom
state in each lattice site. Finally, the
lattice potential is adiabatically lowered to the desired value
such that an unconventional pairing phase with inter-site coherence is
generated during such ``melting'' process. When bosons with
light effective mass are used, they will remian in
the superfluid phase and mediate the attractive
interaction between fermions during an entire procedure.
We point out that this method can be extermely efficient since
the initial molecule condensate ensures that
all final Cooper pairs are close to the zero momentum states. 

The exotic many-body phases can be detected by various approaches. First
of all, CDW order can be observed in a standard time-of-flight measurement
of {\it bosons}, which produces additional peaks
at wavevector, $\bfQ=(\pi,\pi)$.
The spin density correlation and the pair correlation can be further
investigated by studying the noise correlation in a
time-of-flight measurements \cite{ehud}:
the spectrum of pairing phase will have a peak
at zero momentum due to the condensation of Cooper pairs, while
it will peak at momentum ${\bf Q}$ in the SDW phase due to nesting.
Furthermore, one can use Bragg scattering spectroscopy to probe the
gap symmetry of the fermion pairing as proposed in Ref.
[\onlinecite{walter}].
For example, for $d$-wave pairing phase
a zero energy excitation should be observable when
the momentum transfer of the two scattering photons is tuned to match
any two of the nodal points on the Fermi surface.
One can also use rf-spectroscopy to measure the binding energy of
fermion-pairing \cite{chin} and the photoassociation method to
measure the superfluid fraction of the fermion pairs
\cite{photoassociation}.
These techniques are presently used by a number of experimental groups.

In summary we investigated the many-body phase diagrams 
of a boson-fermion mixture in 2D optical lattice. 
For a realistic $^{40}$K-$^{23}$Na or $^{40}$K-$^{87}$Rb system, 
a nature of superfluidity can be controlled by appropriate tuning of the 
atomic interaction interaction strength and optical lattice parameters. 
Our results indicate that experimental studies of 
boson-fermion mixtures of cold atoms could have important implications 
for understanding  the physics of unconventional superconducting materials.

We thank J. Fortagh, W. Hofstetter, L. Mathey, D.S. Petrov, and 
D. Podolsky for valuable discussions.
This work was supported by the NSF (grants DMR-01328074, PHY-0134776),
the Sloan and the Packard Foundations, and by Harvard-MIT CUA.



\begin{thebibliography}{99}
\bibitem{note} We note that while there is no true long-range order
in 2D system at finite temperature due to thermal phase fluctuations,
we expect the correlation length to increase rapidly below the 
mean-field $T_c$ and becomes comparable to the
system size for any realistic system.
\bibitem{scattering_constant}
The background $s$-wave scattering lengths used in our calculation is
given by $a_{K-K}=174$, $a_{Rb-Rb}=100$, $a_{K-Rb}=-330$, 
$a_{Na-Na}=52$ and $a_{K-Na}=-129$ $a_0$, where $a_0$ is atomic unit.
\end{thebibliography}
\end{document}